\documentclass{article}
\usepackage{spconf,amsmath,graphicx,hyperref,xcolor}
\usepackage{subcaption, booktabs, framtex}

\usepackage[font=footnotesize]{caption} 
\usepackage[withpage, printonlyused,nohyperlinks]{acronym}

\title{Explicit modelling of subject dependency in BCI decoding}

\name{Michele Romani$\,^{1,3}$, Francesco Paissan$\,^{1}$, Andrea Fossà$\,^{2,3}$, Elisabetta Farella$\,^{3}$\thanks{This work was done while A. Fossà was an intern at FBK.}}
\address{$^{1}$ University of Trento,  $^{2}$ University of Bologna,  $^{3}$ Fondazione Bruno Kessler}

\begin{document}

\ninept
\maketitle

\acrodef{hci}[HCI]{Human-Computer Interaction}
\acrodef{geq}[GEQ]{Game Experience Questionnaire}
\acrodef{dsp}[DSP]{Digital Signal Processing}
\acrodef{adhd}[ADHD]{Attention Deficit Hyperactivity Disorder}

\acrodef{xr}[XR]{Extended Reality}
\acrodef{mr}[MR]{Mixed Reality}
\acrodef{ar}[AR]{Augmented Reality}
\acrodef{vr}[VR]{Virtual Reality}

\acrodef{bcmi}[BCMI]{Brain-Computer Music Interface}
\acrodef{bci}[BCI]{Brain-Computer Interface}
\acrodef{mi}[MI]{Motor Imagery}
\acrodef{erp}[ERP]{Event-Related Potential}
\acrodef{errp}[ErrP]{Error-Related Potentials}
\acrodef{eeg}[EEG]{Electroencephalography}
\acrodef{lda}[LDA]{Linear Discriminant Analysis}
\acrodef{svm}[SVM]{Support Vector Machine}
\acrodef{vep}[VEP]{Visually Evoked Potentials}
\acrodef{aerp}[AERP]{Auditory Event-Related Potential}
\acrodef{ssvep}[SSVEP]{Steady-State Visually Evoked Potentials}
\acrodef{tvep}[TVEP]{Time-Modulated Visually Evoked Potentials}
\acrodef{mvep}[MVEP]{Motion-Onset Visually Evoked Potentials}
\acrodef{cvep}[CVEP]{Code-Modulated Visually Evoked Potentials}
\acrodef{fft}[FFT]{Fast Fourier Transform}
\acrodef{psd}[PSD]{Power Spectral Density}
\acrodef{csp}[CSP]{Common Spatial Patterns}
\acrodef{ers}[ERS]{Event-Related Synchronization}
\acrodef{erd}[ERD]{Event-Related Desynchronization}
\acrodef{snr}[SNR]{Signal-to-Noise Ratio}
\acrodef{mdrm}[MDRM]{Minimum Distance to Riemannian Mean}
\acrodef{smr}[SMR]{Sensory Motor Rhythms}
\acrodef{itr}[ITR]{Information Transfer Rate}

\acrodef{ai}[AI]{Artificial Intelligence}
\acrodef{dl}[DL]{Deep Learning}
\acrodef{dnn}[DNN]{Deep Neural Network}
\acrodef{cnn}[CNN]{Convolutional Neural Network}
\acrodef{bce}[BCE]{Binary Cross-Entropy}
\acrodef{loso}[LOSO]{Leave-One-Subject-Out}
\acrodef{film}[FiLM]{Feature-wise Linear Modulation}

\acrodef{ml}[ML]{Machine Learning}
\acrodef{mcc}[MCC]{Matthews Correlation Coefficient}
\acrodef{tpe}[TPE]{Tree-Structured Parzen Estimator}

\begin{abstract}

Brain–Computer Interfaces (BCIs) suffer from high inter-subject variability and limited labeled data, often requiring lengthy calibration phases. In this work, we present an end-to-end approach that explicitly models the subject dependency using lightweight convolutional neural networks (CNNs) conditioned on the subject's identity. Our method integrates hyperparameter optimization strategies that prioritize class imbalance and evaluates two conditioning mechanisms to adapt pre-trained models to unseen subjects with minimal calibration data. We benchmark three lightweight architectures on a time-modulated Event-Related Potentials (ERP) classification task, providing interpretable evaluation metrics and explainable visualizations of the learned representations. Results demonstrate improved generalization and data-efficient calibration, highlighting the scalability and practicality of subject-adaptive BCIs.

\end{abstract}


\section{Introduction}
Modern \ac{bci} systems still struggle to generalize and adapt to unseen users, often requiring a subject-specific calibration phase to achieve optimal performance. This difficulty arises from high variability between users, differences in electrode placement and setup, sensor sensitivity, as well as endogenous temporal changes within the same individual, such as accumulating fatigue or physiological fluctuations in brain activity and connectivity across sessions~\cite{10.3389/fnins.2021.733546, ma_large_2022, huang_discrepancy_2023}.
This combination of high inter-subject variability and limited availability of large, labeled datasets further constrains the ability of deep models to generalize, reinforcing the continued dominance of more data-efficient classical approaches~\cite{lotte_review_2018, chevallier2024largesteegbasedbcireproducibility}.
In benchmarks spanning over various architectures and methods, \acp{dnn} rarely surpass the performance of state-of-the-art methods, which remain preferred by the \ac{bci} community for online applications. This is particularly evident for classic subject-dependent machine learning approaches that require calibration, but also for transfer-learning settings, where methods based on Riemannian geometry outperform all others~\cite{Congedo03072017, chevallier2024largesteegbasedbcireproducibility}. 
These methods, in fact,  respect the SPD manifold structure of \ac{eeg} covariance features, offering strong cross-subject generalization.
More recently, however, a number of approaches have emerged that explicitly condition \acp{dnn} on subject-specific features. Nemes et al.~\cite{gyula_nemes_subject_2024}, for instance, proposed an attentive subject-fusion framework that encodes subject information using power spectral density descriptors for motor-imagery \acp{bci}.
Similarly, Sun et al.~\cite{sun_eeg_2024,sun_electroencephalography_2025} introduced a method that integrates conditional identification information to leverage interactions between \ac{eeg} signals and individual traits, thereby enhancing the model’s internal representations.
Originating in the computer-vision domain, \ac{film}~\cite{perez_film_2017} has also proved to be a robust, general-purpose mechanism for selectively manipulating a network’s intermediate features, making it a natural candidate for subject conditioning in \ac{bci} applications.

In this work, we propose an end-to-end approach for modeling subject dependency using lightweight \acp{cnn}, motivated by the high inter-subject variability and scarcity of labeled data typical of \ac{bci} research. We compare two strategies to condition pre-trained models for unseen subjects, starting with minimal calibration data and mimicking the setup steps of a real-time \ac{bci} based on time-modulated \acp{erp}~\cite{5294934}.  
Our main contributions are:
\begin{itemize}
    \item We present a robust and scalable end-to-end approach for conditioning deep neural networks in \ac{eeg} classification tasks;
    \item We introduce a comprehensive hyperparameter optimization strategy that addresses class imbalance issues while ensuring fair comparison with state-of-the-art pipelines;
    \item We propose an incremental fine-tuning methodology that minimizes the amount of subject-specific data required for \ac{bci} calibration.
\end{itemize}

The paper is structured as follows: \cref{sec:method} presents the methodology proposed in this paper, \cref{sec:experiments} describes the experimental setup, and \cref{sec:results} summarizes our findings.

\begin{figure*}[htbp]
    \centering
    \resizebox{.9\linewidth}{!}{
    \begin{subfigure}{0.49\textwidth}
        \centering
        \includegraphics[width=\textwidth]{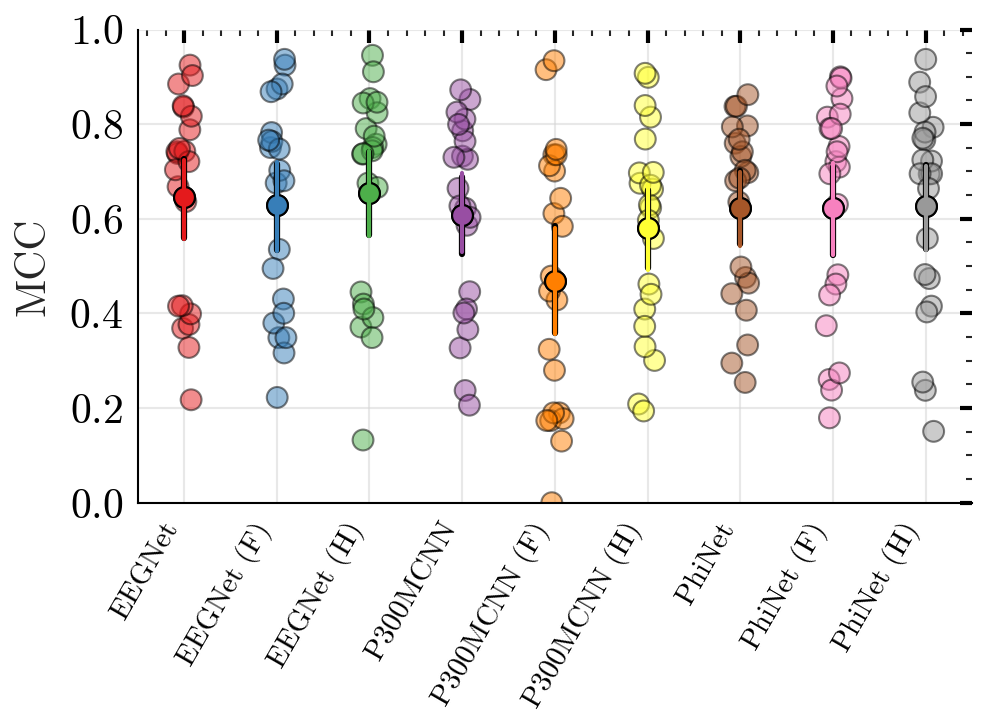}
        \caption{Pre-training phase}
        \label{fig:pretrain}
    \end{subfigure}
    \hfill
    \begin{subfigure}{0.49\textwidth}
        \centering
        \includegraphics[width=\textwidth]{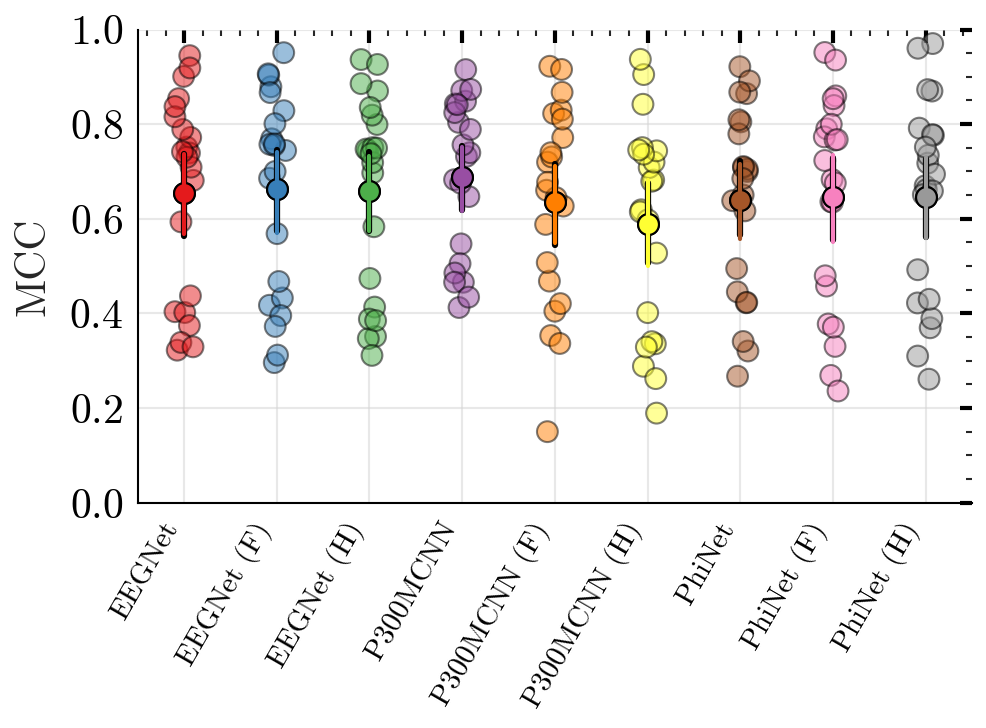}
        \caption{Fine-Tuning Stage (4 batches)}
        \label{fig:finetune}
    \end{subfigure}
    }
    \caption{\textbf{Comparison of MCC Scores.} We analyze how various conditioning strategies affect the performance of the test neural architectures across different training stages.}
    \label{fig:comparison}
    \vspace{-15pt}
\end{figure*}

\section{Method}\label{sec:method}

Given a dataset comprising $N$ subjects $\mathcal{S}\defeq\{s_i\}_{i=1}^N$, each identified by a unique subject-specific identifier $s_i$, we propose two conditioning approaches to explicitly model subject-specific features in neural network inference, rather than treating such features as implicit covariates. We consider a feature extractor $\Psi_\psi: \mathcal{E} \to \mathcal{H} \subseteq \mathbb{R}^d$ that maps EEG signals to latent representations, and a classifier $\Theta_\theta: \mathcal{H} \to \{0,1\}$.
In contrast to previous approaches that either inject handcrafted subject descriptors or rely on auxiliary statistical representations, our methodology directly learns a subject embedding table $\bm{e} \in \mathbb{R}^{N \times d}$ jointly with the parameters of both the feature extractor $\psi$ and classifier $\theta$. Specifically, we augment the feature extraction process $\Psi$ to incorporate subject-specific information through two distinct mechanisms: a projection-based approach and a Feature-wise Linear Modulation (FiLM) layer. This design enables a systematic comparison of how different conditioning strategies influence cross-subject generalization performance and calibration efficiency.

\textit{Approach I -- Projection.} Let $\bm{h} \in \mathbb{R}^d$ denote the features extracted from $\Psi$. This conditioning strategy performs subject-specific modulation in the feature space by computing the projection
\begin{align}
    \widetilde{\bm{h}} = \text{proj}_{\bm{h}} \bm{e}_{s_i} = \parens*{\bm{h}^\top\bm{e}_{s_i}}\,\bm{h},
\end{align}
where $\|\bm{e}_{s_i}\|_2 = 1$ and $s_i \in \mathcal{S}$ denotes the unit-normalized embedding vector associated with the subject that generated the input data. This operation preserves the direction of $\bm{h}$ while scaling its magnitude according to the cosine similarity with the subject embedding $\bm{e}_{s_i}$, thereby introducing adaptive subject-specific amplification or attenuation of the feature response.
Notably, this projection-based conditioning exhibits an interesting geometric property: it effectively learns a subject-specific \textit{receptive field} in the feature space, where features aligned with the subject's learned direction receive amplification, while orthogonal features are suppressed.

\textit{Approach II -- FiLM.} The second conditioning strategy employs Feature-wise Linear Modulation (FiLM) to perform affine transformations on the extracted features. Given the subject embedding $\bm{e}_{s_i} \in \mathbb{R}^{2d}$, we partition it into two halves to obtain the modulation parameters $\bm{\gamma}_{s_i}, \bm{\beta}_{s_i}$, which are later normalized to unit norm.
The conditioned features are then computed as:
\begin{align}
    \widetilde{\bm{h}} = \bm{\gamma}_{s_i} \odot \bm{h} + \bm{\beta}_{s_i},
\end{align}
where $\odot$ denotes element-wise multiplication. This approach provides greater modelling flexibility compared to the projection method, as it allows for both multiplicative scaling ($\gamma$) and additive bias ($\beta$) adjustments on a per-feature basis. Interestingly, while the projection method operates through a single scalar modulation applied uniformly across all feature dimensions, FiLM enables \textit{heterogeneous} feature modulation, where each dimension can be independently scaled and shifted according to subject-specific learned parameters.

\section{Experiments}\label{sec:experiments}

\subsection{Dataset}
The \textit{BrainForm} dataset used in this study was collected through a game-based protocol featuring up to 10 concurrent stimuli across two different tasks, with a total of 22 subjects completing the experiment. Each subject participated in two sessions, each comprising a calibration phase of 60 trials and a total of four runs, including a tutorial. An additional session, including a calibration phase and a free-play run, was completed by 16 subjects. In this study, all optional calibration sessions were included to enrich the training set but excluded from the leave-out fold to ensure a fair comparison of the test results. The review paper on \textit{BrainForm} is currently under revision, after which the dataset will be publicly available.
The dataset was recorded using a g.tec Unicorn system with conductive gel, featuring eight electrodes and a sampling frequency of \SI{250}{\hertz}. Calibration phases from each session were extracted and preprocessed using the \texttt{MNE} toolbox~\cite{GramfortEtAl2013a}. Notch filters at \SI{50}{\hertz} and \SI{60}{\hertz} were applied to remove power-line interference, followed by a band-pass filter between \SI{2}{\hertz} and \SI{15}{\hertz} to restrict the network to the frequency range relevant for \acp{erp}. Data were epoched (baseline -100ms - 0ms) using the trigger-channel events, producing 600 epochs per session (60 Target, 540 Non-Target). Epochs were resampled to \SI{125}{\hertz} for computational efficiency.  Depending on the optimization process (see \cref{sec:optimization}), we scaled the data using a channel-wise standard scaler or robust scaler.

\subsection{Training Setup}
The data were split following a \ac{loso} protocol. For each fold, the training data were further partitioned into training and validation sets with a \num{0.92}:\num{0.08} ratio. The two sessions of the held-out subject were then combined and temporally divided into a fine-tuning set and a test set. This temporal split counterbalanced potential fatigue effects by taking half of each session and merging it with the opposite half, while preserving the original class-label distribution. A batch size of \num{120} was used for the training and validation sets, whereas the fine-tuning and test sets used a batch size of \num{60}, resulting in exactly \num{10} batches each. 
The experiment was divided into two stages: pre-training and incremental Fine-Tuning.

During pre-training, the models were trained on the training set and evaluated on the validation set. We employed an early-stopping mechanism to mitigate overfitting, targeting the validation \ac{mcc} with a patience of 10 epochs. Simultaneously, the learning rate was reduced by a factor of 10 every 20 epochs. At the end of each training, we keep the model weights for the checkpoint with the best validation \ac{mcc}.
During fine-tuning, we start with one batch from the fine-tuning set and progressively increase up to four batches. Each model was fine-tuned and cross-validated using all possible permutations of the selected batches. During the Fine-Tuning stage, we freeze all parameters except for the final classification layer in models without conditioning mechanisms; for conditioned models, we update only the conditioning layer. The learning rate was heuristically standardized to $\num{5e-4}$ for all configurations. The patience of the early-stopping mechanism was reduced to 5 epochs, and as in pre-training, the learning rate was lowered by a factor of 10 every 20 epochs.

\subsection{Balancing Strategies}
Similarly to the P300 paradigm, time-modulated \acp{erp} relies on a Target/Non-Target setup in which Target events occur far less frequently, especially when the \ac{bci} application involves many possible targets. This naturally produces highly unbalanced datasets; in our case, the calibration phase exhibited a label distribution of one Target (T) for every nine Non-Targets (NT).  
To address this imbalance, we focused on using optimization objectives that inherently account for class imbalance, rather than oversampling, as oversampling small classes tends to increase the risk of overfitting. Specifically, we evaluated three approaches employing \ac{bce} loss~\cite{Topsøe2001} and Focal Loss~\cite{lin_focal_2018}:  (i) weighted \ac{bce} loss combined with under-sampling of the NT class, (ii) weighted \ac{bce} loss only, and (iii) Focal Loss only. These strategies were chosen to minimize calibration time while improving generalization to unseen subjects, reflecting the constraints of real-world online \acp{bci}.

\begin{figure}[t]

\begin{minipage}[b]{1.0\linewidth}
  \centering
  \centerline{\includegraphics[width=\textwidth]{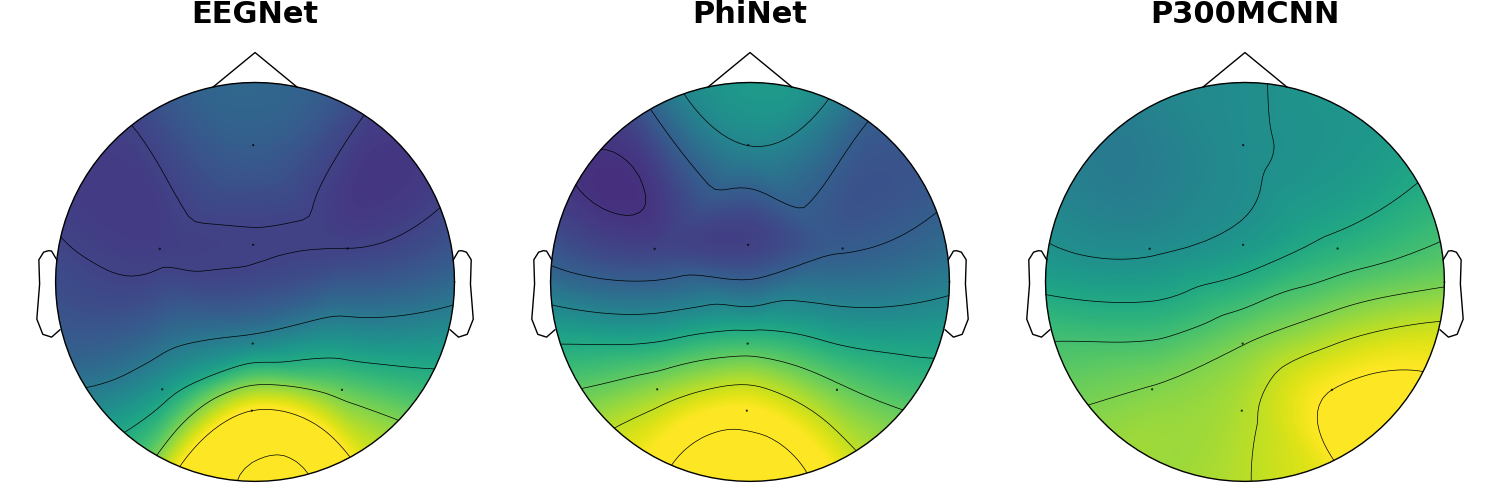}}
  \centerline{A}\medskip
\end{minipage}
\begin{minipage}[b]{.32\linewidth}
  \centering
  \centerline{\includegraphics[width=\linewidth]{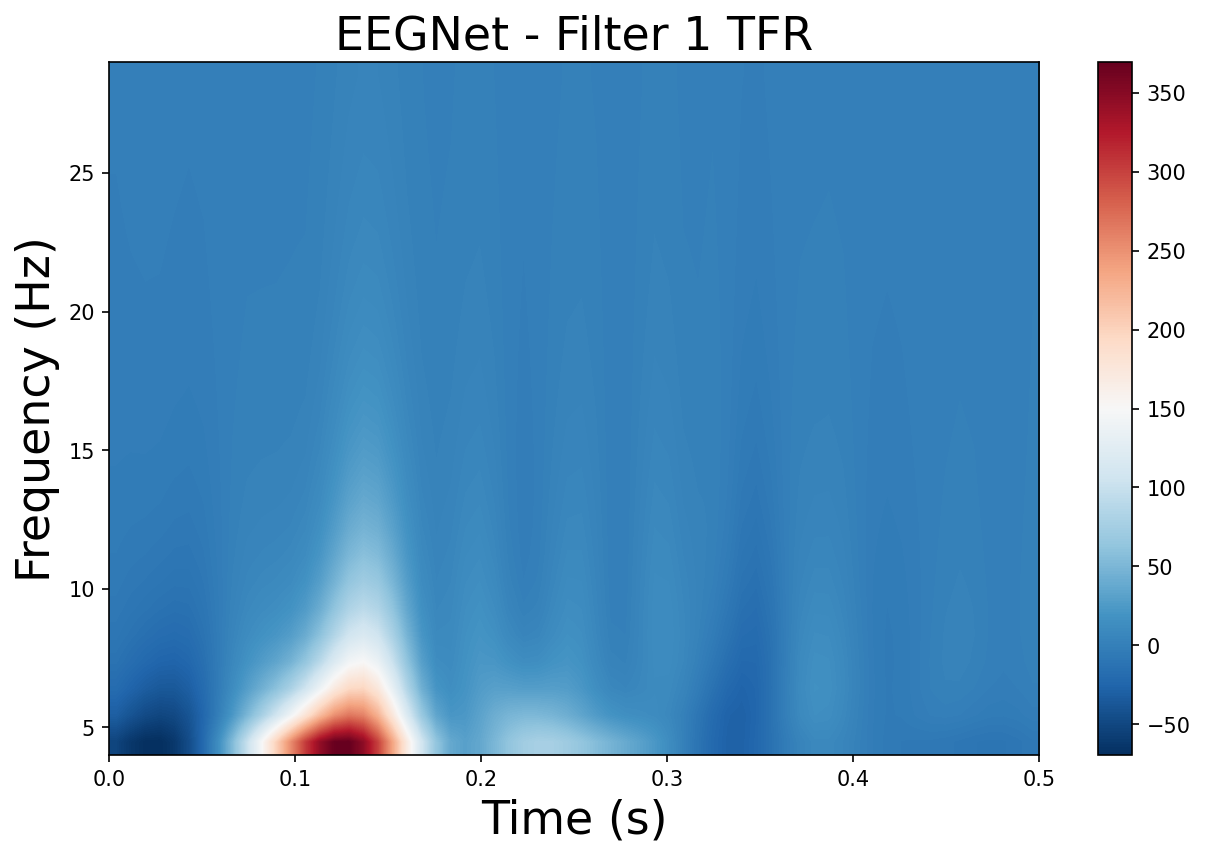}}
  \centerline{B1) EEGNet}\medskip
\end{minipage}
\hfill
\begin{minipage}[b]{0.32\linewidth}
  \centering
  \centerline{\includegraphics[width=\linewidth]{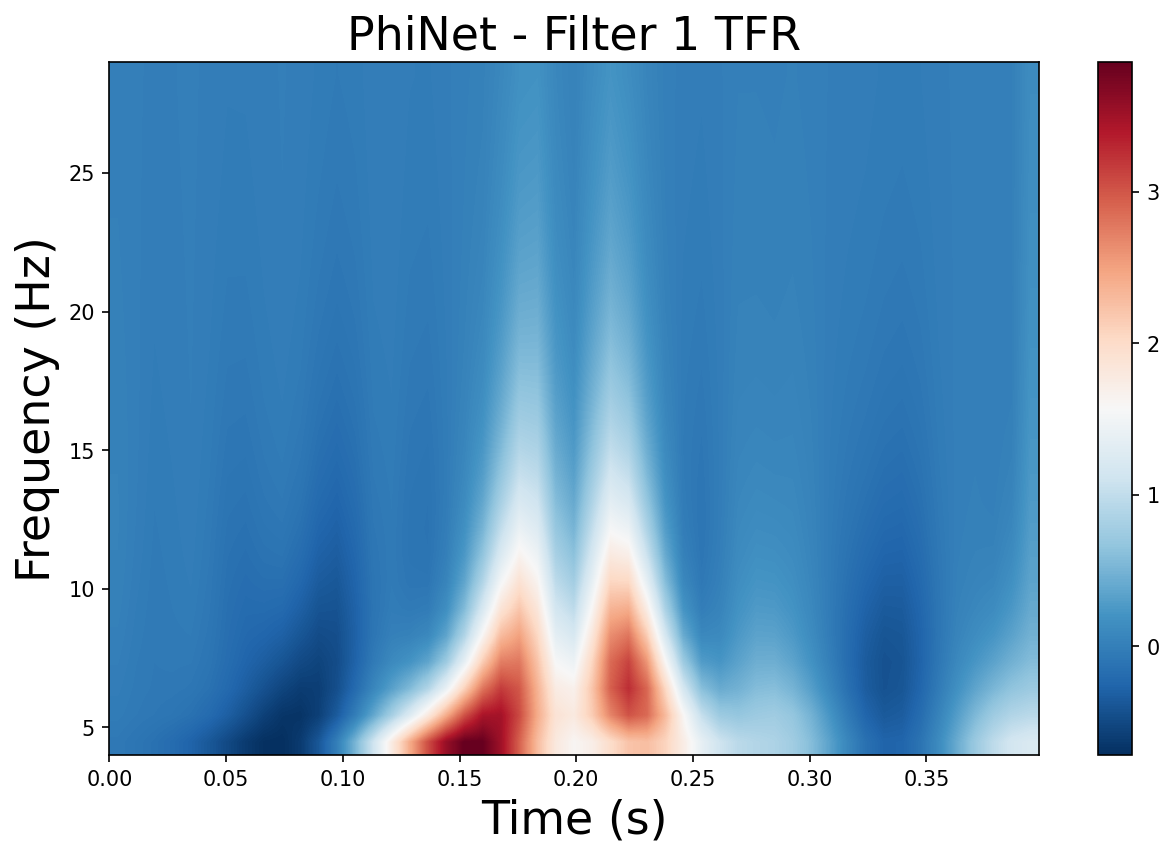}}
  \centerline{B2) PhiNet}\medskip
\end{minipage}
\hfill
\begin{minipage}[b]{0.32\linewidth}
  \centering
  \centerline{\includegraphics[width=\linewidth]{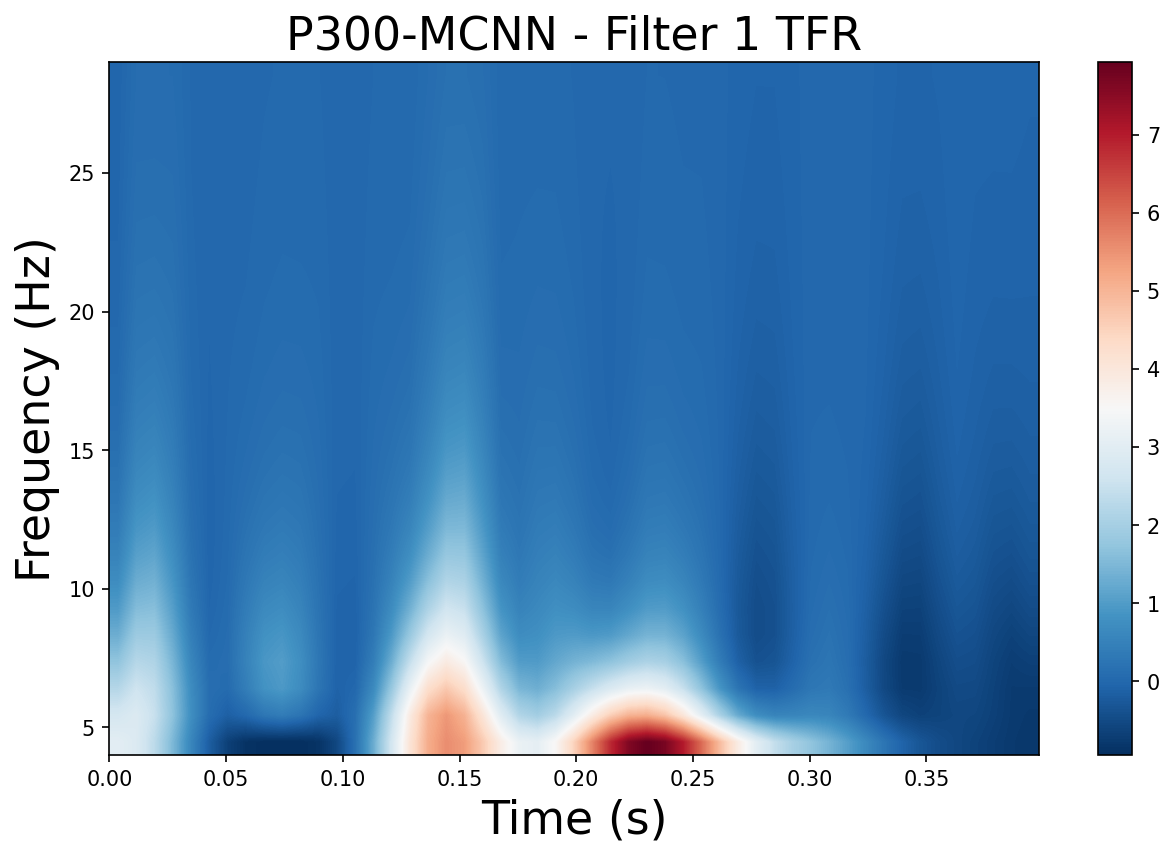}}
  \centerline{B3) P300MCNN}\medskip
\end{minipage}

\caption{\textbf{Channel Relevance and T--NT Filter Response Difference.} A) Weight energy distribution for the three models. Channel importance increases from cool to warm colors.  
B) Difference in filter responses between target and non-target stimuli, displayed as a function of frequency and time. Warm colors indicate stronger responses for targets (T), cool colors indicate stronger responses for non-targets (NT).}
\vspace{-16pt}
\label{fig:res}
\end{figure}
\subsection{Evaluation Metrics}
Since \ac{bci} applications are aimed at support systems, it is critical to acknowledge both false negatives and positives when training a model. In the specific case of a \ac{erp}-based \ac{bci}, the former correspond to the system non-responding and the latter to the system activating the wrong command.
Given the complexity of \ac{eeg} classification and the often unbalanced distribution of training labels, metrics like \textit{accuracy} and \textit{F1-score} do not sufficiently capture the model's ability to discriminate between classes. We opted for \ac{mcc}~\cite{10.1093/bioinformatics/16.5.412}, a robust method that is reportedly more reliable for two-class confusion matrix evaluation~\cite{chicco_matthews_2021}. Thus, \ac{mcc} was selected as the target metric in all optimization steps.
In line with the standards of MOABB~\cite{chevallier2024largesteegbasedbcireproducibility} for the binary P300 task, a well-known benchmark in the \ac{bci} community, we also report the ROC AUC score on our companion website. %

\subsection{Feature Extractors}
The following architectures were selected for their relatively low parameter count and their reliance on depthwise–separable convolutions, which have proven effective for learning spatial filters and extracting temporal summaries from EEG feature maps. Specifically, we considered EEGNet \cite{lawhern_eegnet_2018}, a widely used architecture in the \ac{bci} field inspired by the MobileNets family\cite{howard_mobilenets_2017}; P300MCNN \cite{liu_best_2024}, a \ac{cnn} specialized for P300 classification; and PhiNet \cite{paissan_phinets_2021}, a lightweight and scalable model that integrates squeeze-and-excitation blocks\cite{hu_squeeze-and-excitation_2019} to enhance feature attention. 
This choice is motivated by their suitability for embedded \ac{bci} deployment, where computational and energy resources are constrained, making scalability and real-time usability key requirements.

\subsection{Hyper-parameters Optimization}\label{sec:optimization}

The hyperparameters for each architecture were optimized using Optuna~\cite{optuna_2019} with \ac{tpe} and validation \ac{mcc} as objective. Similar to \cite{BORRA2024109097}, we first conducted a \ac{loso} search to optimize the parameters across all subjects. We used these trials to find the median subject, identified by computing the absolute distance from the global mean using \ac{mcc} across all Optuna trials. Finally, we conducted the hyperparameter optimization for all architectures using only the median subject as the held-out subject to speed up the search process. To ensure robustness and to account for variability due to random initialization~\cite{DBLP:journals/corr/abs-2109-08203}, all analyses were repeated with multiple random seeds, yielding an average standard deviation of \num{0.0159} for our metric after five repetitions.

Among the optimized parameters, the most relevant included kernel length, time window, scaling method, and loss function. The Optuna search yielded the following best-performing configurations: EEGNet combined with RobustScaler, Focal Loss, and a 0.5s window; P300MCNN with RobustScaler, Weighted BCE Loss, and a 0.6s window; and PhiNet with RobustScaler, Focal Loss, and a 0.35s window. Each configuration was evaluated with no conditioning layer, with \ac{film} conditioning, and with \textit{H projection} conditioning. The final parameter counts satisfied our efficiency requirements, with EEGNet requiring approximately \num{4500} parameters, P300MCNN about \num{9500}, and PhiNet around \num{3500}.

\begin{figure}[t]
    \centering
    \includegraphics[width=0.9\linewidth]{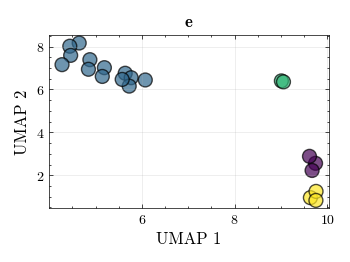}
    \vspace{-8pt}
    \caption{\textbf{Subject-Dependent Clustering.} UMAP projection of the subject embedding $\bm{e}$ table extracted from PhiNet (F) after pretraining on all subjects except \texttt{EXP\_P12}. Each color represents a cluster extracted using k-means clustering.}
    \label{fig:umap}
\end{figure}

\begin{table*}[t]
\centering
\resizebox{0.94\textwidth}{!}{%
\begin{tabular}{lccccc}
\toprule
\textbf{Model} & \textbf{Zero-Shot MCC} & \textbf{FT I1 MCC} & \textbf{FT I2 MCC} & \textbf{FT I3 MCC} & \textbf{FT I4 MCC} \\
\midrule
EEGNet & 0.6460 ± 0.2151 & 0.6518 ± 0.2102 & 0.6532 ± 0.2112 & 0.6535 ± 0.2120 & 0.6539 ± 0.2128 \\
EEGNet (H) & \textbf{0.6535 ± 0.2227} & 0.6568 ± 0.2076 & 0.6574 ± 0.2093 & 0.6581 ± 0.2091 & 0.6577 ± 0.2099 \\
EEGNet (F) & 0.6289 ± 0.2252 & \textbf{0.6574 ± 0.2102} & \textbf{0.6609 ± 0.2111} & \textbf{0.6620 ± 0.2120} & \textbf{0.6621 ± 0.2133} \\
\midrule
P300MCNN & \textbf{0.6088 ± 0.2075} & \textbf{0.6404 ± 0.1844} & \textbf{0.6621 ± 0.1737} & \textbf{0.6772 ± 0.1698} & \textbf{0.6890 ± 0.1661} \\
P300MCNN (H) & 0.5799 ± 0.2135 & 0.5790 ± 0.2040 & 0.5838 ± 0.2104 & 0.5871 ± 0.2159 & 0.5888 ± 0.2204 \\
P300MCNN (F) & 0.4687 ± 0.2777 & 0.6064 ± 0.2085 & 0.6308 ± 0.2048 & 0.6324 ± 0.2072 & 0.6349 ± 0.2089 \\
\midrule
PhiNet & 0.6227 ± 0.1906 & 0.6361 ± 0.2019 & 0.6364 ± 0.2006 & 0.6377 ± 0.1996 & 0.6394 ± 0.1980 \\
PhiNet (H) & \textbf{0.6269 ± 0.2229} & \textbf{0.6389 ± 0.2154} & \textbf{0.6413 ± 0.2121} & \textbf{0.6439 ± 0.2101} & 0.\textbf{6469 ± 0.2085} \\
PhiNet (F) & 0.6231 ± 0.2388 & 0.6359 ± 0.2275 & 0.6400 ± 0.2257 & 0.6431 ± 0.2221 & 0.6460 ± 0.2206 \\
\bottomrule
\end{tabular}%
}
\caption{Zero-shot and fine-tuning performance. \textbf{Bold} characters denote the best score of each column. Models denoted with H use the projection-based conditioning, while F use FiLM-based conditioning.}
\label{tbl:results}
\vspace{-14pt}
\end{table*}

\section{Results}\label{sec:results}

\subsection{Quantitative Results}
The results reported in \cref{tbl:results} show that models equipped with a conditioning layer generally achieve better or comparable generalization during the pre-training phase, with the exception of the P300MCNN architecture. More specifically, our proposed \textit{H projection} method attains a slightly higher score than \ac{film} for PhiNet, clearly outperforms it in the EEGNet and P300MCNN models. During the Fine-Tuning phase, the models tend to reach a noticeable performance plateau after the first fine-tuning step, with subsequent batches yielding only minor improvements in performance. \ac{film} reaches better performance in the P300MCNN (0.6268 against 0.5948) and obtains the best score among EEGNet architectures with 0.6621. In the PhiNet architectures, the \textit{H projection} method scores the best performance. Additional results using ROC AUC and Balanced Accuracy are reported on our companion website\footnote{
\url{https://bromans.github.io/icassp26-companion/}}.

\subsection{Discussion}
Compared to similar fine-tuning experiments on \ac{erp} classification, we ensured that the length of the test set would remain constant across the different training stages, ensuring maximum comparability between configurations.
The held-out subjects in each fold exhibited substantial variability, as illustrated in \cref{fig:comparison}. This variability decreased somewhat after the first fine-tuning step, indicating improvement for most subjects, but then remained relatively stable despite the addition of more batches. This behaviour likely reflects the performance plateau observed in the models and may also indicate that the available data are insufficient for the models to fully generalize to the task.

P300MCNN achieved its best performance only after the fine-tuning phase and was the only architecture paired with weighted \ac{bce}. PhiNet maintained comparable performance despite using a very short time window of 0.35 s, which is barely sufficient to capture \ac{erp} components. Focal loss emerged as the preferred method for addressing class imbalance, which aligns with expectations since it dynamically adjusts to focus on hard-to-classify cases. The robust scaler, which uses the median and interquartile range rather than the mean and standard deviation, also contributed to reducing the impact of outliers and artifacts.

\subsection{Explainability}
We performed an analysis of the trained models for \ac{eeg}, comparing the three architectures to quantify the contribution of each channel to the model’s decisions. Specifically, we analyzed the weights of the first convolutional layer, which is closest to the raw signal and thus provides a direct indication of channel relevance. The contribution of each channel was computed as the energy of its weights (sum of squares), offering a measure of the overall channel’s influence: $
I_c = \sum_{d,k} w_{c,d,k}^2$, where $d$ represents the number of filters in the convolution.
This measure was normalized and compared across multiple models and checkpoints to produce topographic maps, statistical distributions, and channel rankings. As shown in \cref{fig:res}A, and consistent with findings in the literature, the analysis highlights both the most informative channels for each model and those that retain a stable role across architectures, thereby providing an objective measure of cross-model consistency and interpretability.

To complement the weight-based analysis, we examined the activations of the same convolutional layers analyzed previously to characterize the time–frequency patterns emphasized by each architecture in the early stages of processing. Specifically, we computed the filter responses to EEG trials under both target and non-target conditions. These activations were then transformed into the time–frequency domain using Morlet wavelets, yielding spectrograms (\cref{fig:res}B) for each filter that capture the oscillatory dynamics associated with the input signals.

After pre-training, we extracted the weights of the embedding table from the conditioning layers and applied UMAP~\cite{mcinnes2020umapuniformmanifoldapproximation} to project them into a lower-dimensional space for visualization. As shown in Fig.\ref{fig:umap}, subjects form distinct clusters, likely reflecting individual differences in their \ac{eeg} signals and \ac{erp} components.
These observable weight clusters suggest that similarities between subjects could be exploited to improve both the initialization of the embedding table and the training process.

\section{Conclusion}
We present an end-to-end strategy for modeling subject dependency through latent representation conditioning. The proposed method demonstrates architecture-agnostic applicability, scalability, and data efficiency across three benchmarked architectures. Hyperparameter optimization addressing class imbalance, coupled with interpretable metrics and explainable visualizations, provides insights into model learning dynamics. Future work will extend this conditioning framework to diverse datasets and BCI tasks, with validation in online experimental settings.

\bibliographystyle{IEEEbib}
\bibliography{ICASSP25}

\end{document}